\documentstyle[epsfig]{aipproc}

\begin{document}
\title{Confronting Synchrotron Shock and Inverse
Comptonization Models with GRB Spectral Evolution}

\author{A. Crider$^*$, E.P. Liang$^*$, and R.D. Preece$^{\dagger}$}
\address{$^*$Rice University, Houston, TX 77005-1892\\
$^{\dagger}$University of Alabama at Huntsville, Huntsville, AL 35899}

\maketitle

\begin{abstract}
The time-resolved spectra of gamma-ray bursts (GRBs)
remain in conflict with many
proposed models for these events.  After proving that most of the bursts in our
sample show evidence for spectral ``shape-shifting'',  
we discuss what restrictions that BATSE time-resolved burst
spectra place on current models. 
We find that the synchrotron shock model does not allow
for the steep low-energy spectral slope observed in many bursts, including GRB
970111.  
We also determine that saturated Comptonization with
\emph{only} Thomson thinning fails to explain the observed rise and fall of the
low-energy spectral slope seen in GRB 970111 and other bursts.  This implies
that saturated Comptonization models must include some mechanism which can 
cause the Thomson depth to increase intially in pulses.
\end{abstract}

\section*{Introduction}

While observations of gamma-ray burst counterparts 
appear to have determined the distance scale of these
enigmatic events [1], questions regarding 
the emission mechanism remain unanswered.  
Models are tested against the newly discovered X-ray and optical 
power-law tails seen in two GRBs [2,3],
however most of these were flexible enough
to accomodate the observations.  Multiple gamma-ray emission mechanisms
have also been found to be consistent with the \emph{time-integrated}
spectra making this another poor test of GRB models. 

In order to evaluate GRB models, we look to the time evolution of GRB
spectra.  Most spectral evolution studies have looked only at the evolution of 
spectral hardness, which has historically been represented by either 
a hardness ratio or by the peak of the $\nu F \nu$ spectrum, $\rm{E_{peak}}$
[4].
We examine this as well as the evolution of the low-energy asymptotic slope
$\alpha$ (as defined by the Band GRB function [5]). 
In the following section, we show that
most bursts require evolution of this second parameter. 
Thus the shape, as well as
the break energy, of the spectra evolves as a function of time.
Models which were consistent with the time-integrated specta can be 
inconsistent with the time-resolved spectra.  Below we confront
first the synchrotron shock model and then the inverse Comptonization model
with our time-resolved BATSE burst spectra.  We find that both models
are currently unable to explain our observations.

\section{Spectral Shape-Shifting Bursts}

We first demonstrate that time-integrated burst spectra 
do not well represent the time-resolved spectra. 
Previously we have found that a minimum first-order trend in $\alpha$ is shown
to exist in at least 46 of the 79 bursts we analyzed [6]. 
It is possible that some higher order
evolution of $\alpha$ may occur in these bursts.
To test for this, we calculated the $\chi^{2}$ between the time-resolved,
fitted values of $\alpha$ and the weighted average value of $\alpha$
in each burst.  We found that only 30 out of the 79 bursts are consistent
with an invariant $\alpha$, assuming $\rm{Q}(\chi^{2}) > 0.001$ as an acceptable
cutoff [7].  We conclude that the remaining 63\% of the bursts
have a varying $\alpha$, consistent with the results of [6].
Unfortunately, time-resolved $\beta$ values are
poorly determined in a majority of
the bursts in this sample.

\begin{figure} [t] 
\centerline{
\psfig {figure=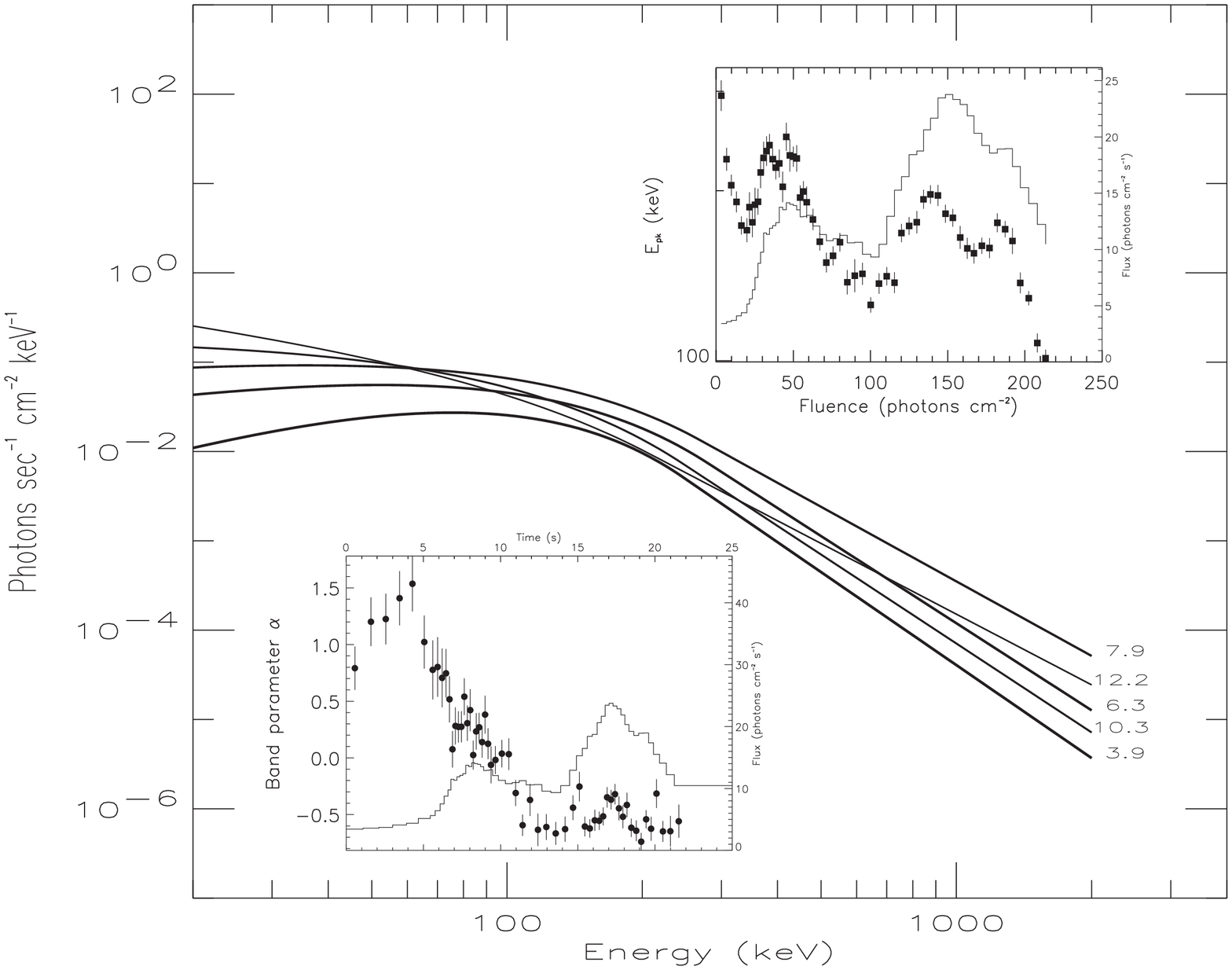,height=2.5in,width=4.0in}}
\vspace{10pt}
\caption
{Evolution of the Band {\it et al.} GRB spectral
function for 3B 970111.
Each line is marked with the
time (in s) corresponding to the beginning of the time bin.
Note that a typical statistical $\beta$ error $\sigma_{\beta} \approx 0.6$.
[Upper inset]
Evolution of $E_{\rm{peak}}$ (squares, logarithmic scale) and photon flux
(histogram, linear scale) with respect to fluence.
[Lower inset] Evolution of $\alpha$ (circles) and photon flux
(histogram) with respect to time.  Error bars represent 1$\sigma$ confidence
level.
The early initial values and evolution of $\alpha$ 
in this burst are difficult to reconcile with the synchrotron shock model.}
\label{fig1_5773}
\end{figure}

As an example of spectral shape-shifting in GRBs, we examine 
the early evolution of GRB 970111 (see Figure \ref{fig1_5773}).
We find the first
several seconds to be remarkably similar to GRB 910927 (see Figure~1 of
[6]).  Both bursts exhibit steep \emph{positive} low-energy
power slopes.  Fitting the Band GRB function [5] to GRB 970111,
we find
$\alpha_{\rm{max}}=+1.5\pm0.2$, a value consistent with the maximum $\alpha$
in GRB 910927. 
We would expect this value of $\alpha$ from 
synchrotron self-absorption.
However, with synchrotron self-absorption, $\alpha$ is either $+\frac{3}{2}$ in
the case of a nonthermal plasma or $+1$ for a thermal plasma [8].  
This mechanism cannot
produce the variation of observed $\alpha$ values.

\section{Synchrotron Shock Model}

One prediction of synchrotron shock models is easily tested with
GRB spectral data.   For these models, the asymptotic low-energy
photon slope $\alpha$ (below the spectral break)
should be between $-\frac{2}{3}$, in the case of an instantaneous
synchrotron spectrum, and $-\frac{3}{2}$, when the spectrum is integrated over
the radiative decay of electron energies [9].  These predictions are
tested by Cohen et al. [9] 
with 11 \emph{time-integrated} GRBs.  In these 11 cases, the slope of
the low-energy spectra is indeed in this range.  However, time-resolved spectra
reported in Crider et al. [6]
clearly show evidence for GRB spectra which cannot be explained with this
mechanism.  Examination of a larger set of bursts show that many bursts have 
maximum $\alpha$ values beyond the ``line-of-death'' ($\alpha > -\frac{2}{3}$)
predicted by synchrotron shock models [10].

\begin{figure} [t] 
\centerline{
\psfig {figure=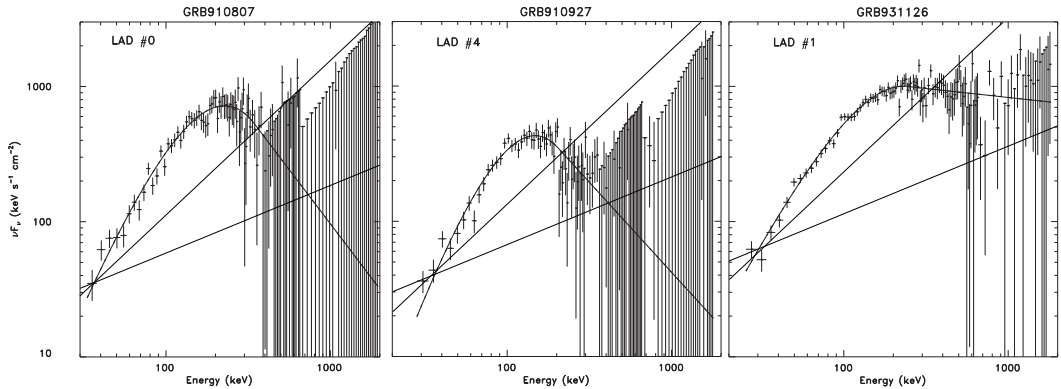,height=2.0in,width=5.5in}}
\vspace{10pt}
\caption{Spectra from GRB 910807, GRB 910927, and GRB 931126 for the time bin in each
where the maximum steepness in the low-energy slope occurred.  The fitted
$\alpha$ for these are, from left to right, $+1.1\pm0.3$, $+1.6\pm0.3$, 
and $+0.4\pm0.1$.  Also plotted are the limits between which
the low-energy asymptotic slope should be for the synchrotron shock model
($-\frac{2}{3} \geq \alpha \geq -\frac{3}{2}$).
The observed slopes are clearly inconsistent with the
synchtron shock predictions.}
\label{vFv}
\end{figure}

To evaluate the synchrotron shock model quantitatively, we began
by finding the probabilities Q of randomly getting poorer $\chi^{2}$ values
when fitting the Band GRB function [5] to time bins selected from GRB
910807, GRB 910927, and GRB 931126.  These are respectively
0.93, 0.50, and 0.25 suggesting that the Band et al. function [5]
adequately describes the observed spectra.
When we instead fixed
$\alpha = -\frac{2}{3}$ to represent the maximum slope allowed
by synchrotron shock models, we found Q values of
$5\times10^{-5}$, $1.5\times10^{-11}$,
and $7 \times 10^{-11}$, which makes this  
model unacceptable in explaining the data. 
Also, when fitting
the time bin in GRB 970111
where $\alpha=+1.5\pm0.2$ gave Q=0.32 for the Band et al. function compared to
Q=$8.5\times10^{-21}$ when $\alpha$ is fixed to $-\frac{2}{3}$.
Three sample spectra from these bursts are plotted in Figure \ref{vFv} in a fashion
similar to that
of Cohen et al. [9] where the limits of the
synchrotron shock model are
overlayed onto the deconvolved burst spectra.  This figure
clearly shows that these bursts' spectral slopes are outside the range 
predicted by the synchrotron shock model.

\section{Saturated Comptonization Model}

\begin{figure} [t] 
\centerline{
\psfig {figure=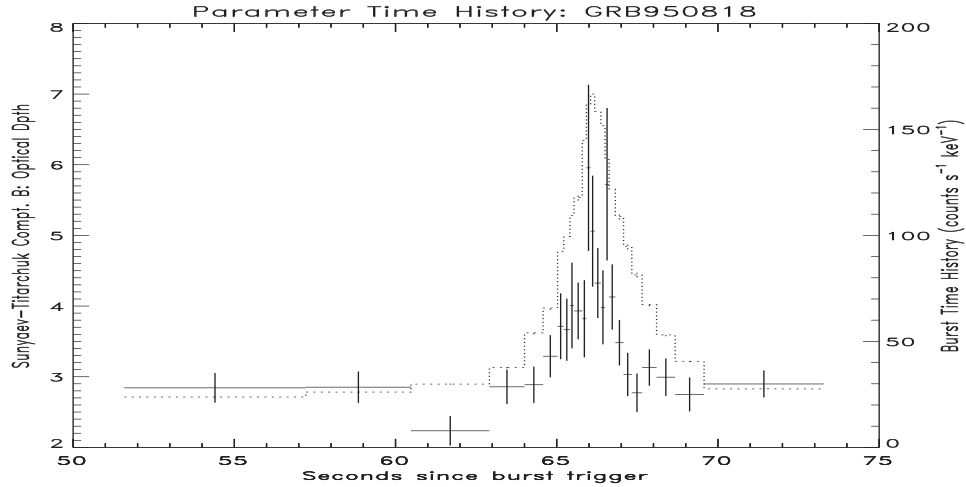,height=2.5in,width=5in}}
\vspace{10pt}
\caption{The evolution of the Sunyaev-Titarchuk Thomson depth
$\tau_{\rm{T}}$ (crosses) and photon flux (histogram)
with time for GRB 950818.  These points are from
fits to the 30-300 keV BATSE LAD spectra.  This evolution of
$\tau_{\rm{T}}$ is difficult to reconcile with a simple Thomson
thinning model.}
\label{taurise_3765}
\end{figure}

The generally decreasing nature of $\alpha$
from values approaching $+2$ is suggestive of saturated inverse
Compton scattering in a plasma experiencing Thomson thinning 
[6,11].
However, many bursts show some evidence for an \emph{increase} in $\alpha$
at the beginning of pulses
which would represent a Thomson thickening.  The observed changes in $\alpha$
cannot be attributed to statistical fluctuations or systematics.
GRB 950818 is an excellent example.  To show more clearly
how the Thomson depth
$\tau_{\rm{T}}$ evolves in this burst, we fit it with the Sunyaev-Titarchuk
model.  The evolution of $\tau$ determined from these fits is shown in 
Figure \ref{taurise_3765}.
While this model does not often fit gamma-ray bursts well,
fits with the Band
GRB function [5] and comparisons with our simulated inverse Compton
spectra [11] indicate that the spectral break is always below
100 keV and that $\tau_{\rm{T}}$ should be $\sim4$.  In this parameter
space, the Sunyaev-Titarchuk model is appropriate if we ignore data above
the spectral break which will be heavily influenced by any non-thermal
component.  The results of fitting this model to our data appear in
the figure above and show that $\tau_{\rm{T}}$
clearly increases and then falls during this burst.
One means by which $\tau_{\rm{T}}$ may rise and fall would be an initial 
compression of the plasma. 
Future research will test if this is a valid hypothesis.

\section{Discussion}

We have found that most bursts show some degree of spectral ``shape-shifting"
making the time-integrated spectrum an inadequate diagnostic
of gamma-ray burst models.  The high value of the low-energy spectral
slope $\alpha\sim+1.5$ seen in some bursts
and the flattening of this slope with time is
incompatible with the synchrotron shock model.  Similarly,  
simple Thomson thinning from an initial state of saturated Comptonization
cannot explain the rise and fall of $\alpha$ seen in some bursts.  
It may be possible to reproduce the observed spectra with a changing
superposition of
self-absorbed synchrotron spectra.  However this has not yet been explored
in detail within the construct of the blast-wave scenario.

\vspace{0.25in}

\noindent AC thanks NASA-MSFC for his GSRP fellowship.  This work is partially
supported by NASA grant NAG5-3824.

\end{document}